\documentclass[prb,preprint]{revtex4-1} 
\usepackage{amsmath}  % needed for \tfrac, \bmatrix, etc.
\usepackage{amsfonts} % needed for bold Greek, Fraktur, and blackboard bold
\usepackage{graphicx} % needed for figures
\usepackage{lineno}
\begin{document}

% Be sure to use the \title, \author, \affiliation, and \abstract macros
% to format your title page.  Don't use lower-level macros to  manually
% adjust the fonts and centering.

\title{Think Inside the Box: Recreating Rutherford’s Scattering}
% In a long title you can use \\ to force a line break at a certain location.

%When submitting the manuscript for review, do not include the author's name or institution
\author{Helio Takai}
\email{htakai@pratt.edu} % optional
\altaffiliation[Pratt Institute ]{200 Willoughby Ave, Brooklyn, USA}
\affiliation{Department of Mathematics and Science, Pratt Institute, Brooklyn, NY 11205}
 % optional second address
% If there were a second author at the same address, we would put another 
% \author{} statement here.  Don't combine multiple authors in a single
% \author statement.
%\affiliation{Department of Physics, Weber State University, Ogden, UT 84408-2508}
% Please provide a full mailing address here.

%\author{David P. Jackson}
%\email{ajp@dickinson.edu}
%\affiliation{Department of Physics, Dickinson College, Carlisle, PA 17013}

% See the REVTeX documentation for more examples of author and affiliation lists.

 \date{\today}

\begin{abstract}
The Rutherford scattering experiment, conceived by Ernest Rutherford and carried out by Hans Geiger and Ernest Marsden, provided the first direct evidence of a compact atomic nucleus, fundamentally altering the understanding of atomic structure. The original experiment involved directing alpha particles from a radioactive source toward a thin gold foil and observing their deflection using a zinc sulfide scintillation screen. This paper presents an accessible and low-cost approach to reproducing the key features of the Rutherford experiment using commercially available components, offering a practical platform for educational demonstrations and introductory nuclear physics investigations.
\end{abstract}
% AJP requires an abstract for all regular article submissions.
% Abstracts are optional for submissions to the "Notes and Discussions" section.

\maketitle % title page is now complete

\section{Introduction} % Section titles are automatically converted to all-caps.
% Section numbering is automatic.

Between 1909 and 1911, Ernest Rutherford and his colleagues Hans Geiger and Ernest Marsden demonstrated that atoms contain a small, dense nucleus \cite{rutherford, geigermarsden}. Their landmark experiment challenged J. J. Thomson’s “plum-pudding” model, which pictured electrons (the “plums”) embedded in a diffuse positive medium (the “pudding”)\cite{thomson}. It was argued at that time that lightweight $\beta$-particles would be scattered by atomic electrons, whereas the much heavier $\alpha$-particles would pass straight through matter with negligible deflection. Contrary to this expectation, Geiger and Marsden observed $\alpha$-particles, emitted by a radioactive source and incident on thin gold foils, scattering through large angles, recorded as scintillations on a zinc-sulfide screen. Such dramatic deflections were possible only if each atom contained a compact, positively charged core. This insight not only revolutionized the atomic model but also established the basic principle of using scattering to probe subatomic structure—a method still central to modern nuclear and particle physics, where high-energy beams collide and sophisticated detectors track the resulting fragments.

Opportunities for students to witness genuine nuclear scattering, however, remain rare. Classroom demonstrations often rely on mechanical analogs, substituting rolling balls for $\alpha$-particles, computer simulations, electro-mechanical emulations, or on legacy laboratory equipment that is neither portable nor widely available\cite{tyson,popenlab,calheiro,eaton}. Practical barriers such as vacuum chambers, radioactive sources, and specialized $\alpha$-particle detectors, reinforce this scarcity. Yet affordable, off-the-shelf technologies and open-source software, now make it feasible to design compelling experiments that bring modern physics into ordinary instructional spaces.

This paper introduces a tabletop nuclear-scattering experiment developed for accessibility and affordability. Using consumer-grade components and readily available materials, the setup lets students detect and analyze $\alpha$-particles, making the otherwise invisible atomic realm tangible. The apparatus is optimized for classroom use and public demonstration, creating an engaging platform for exploring the fundamentals of nuclear scattering.

\section{Experimental Setup}

\begin{figure}[h!]
\centering
\includegraphics[width=5in]{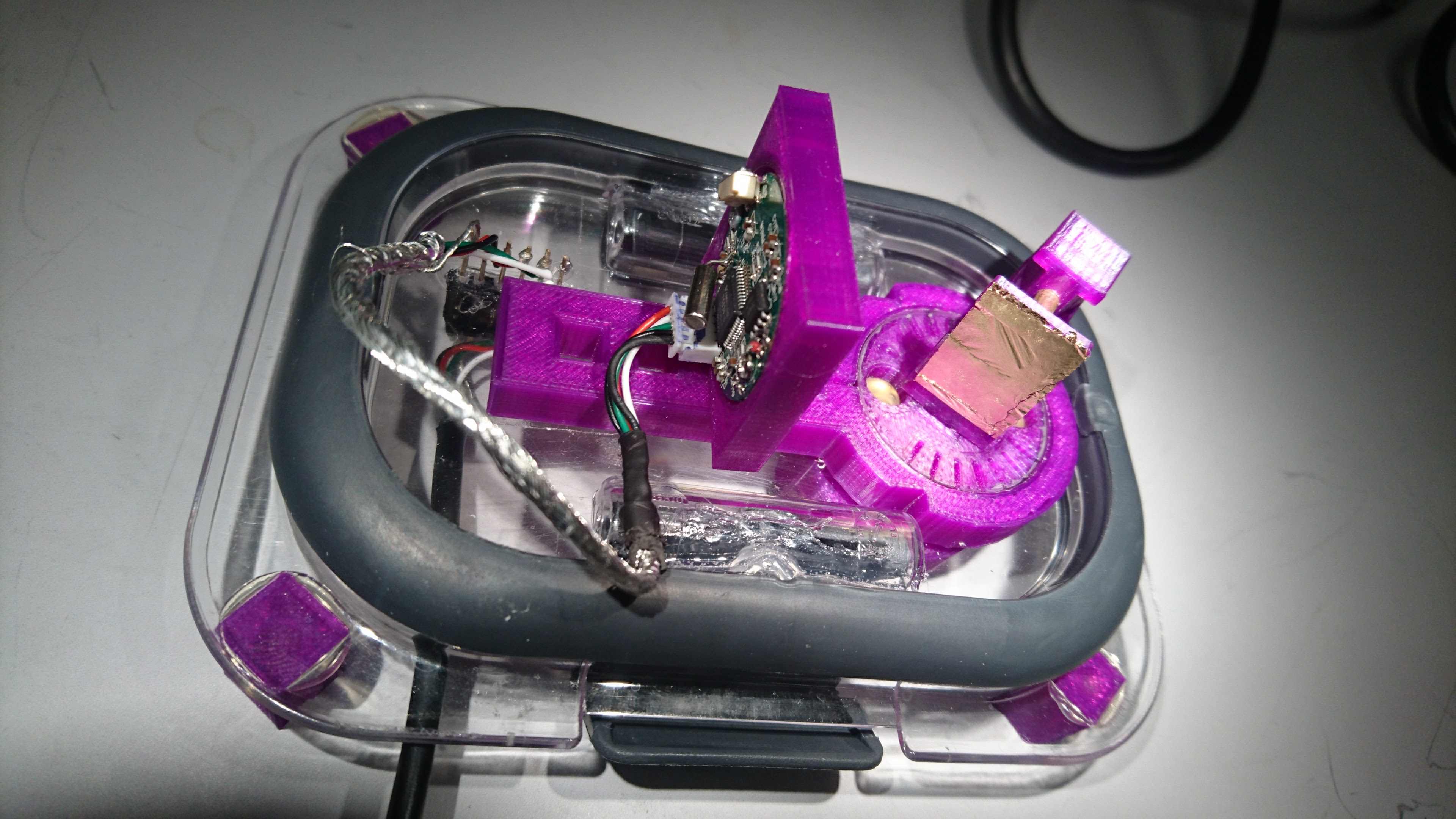}
\caption{The experimental apparatus, featuring the gold foil target and the printed circuit board extracted from a webcam (CCD mounted on the reverse side). The supporting structure is 3D-printed and mounted on the lid of a food container, which serves as the vacuum chamber. }
\label{apparatus1}
\end{figure}

The experimental setup shown in Figure \ref{apparatus1}, introduces the concept of an affordable scattering experiment. All components shown are available as consumer technology commonly found, material from art and kitchen stores, and manufactured using 3D printers. The only part that requires more attention is a radioactive source, that is protected in its own casing. 

\subsection{Particle detector}

Particle detection in nuclear physics experiments typically relies on a range of sophisticated sensors, including semiconductor detectors, plastic scintillators coupled to photomultipliers, and various types of gas detectors. While highly effective, these systems require specialized signal-processing equipment and are generally impractical for classroom use due to their complexity and cost.

To overcome these limitations, this project repurposes a consumer-grade CCD webcam to detect $\alpha$ particles. CCDs function by collecting charge generated through the photoelectric effect in each pixel, which is then digitized to form an image. However, ionizing particles—such as muons from cosmic rays or alpha particles from radioactive sources—can also deposit charge in the sensor, resulting in detectable events.

\begin{figure}[h!]
\centering
\includegraphics[width=5in]{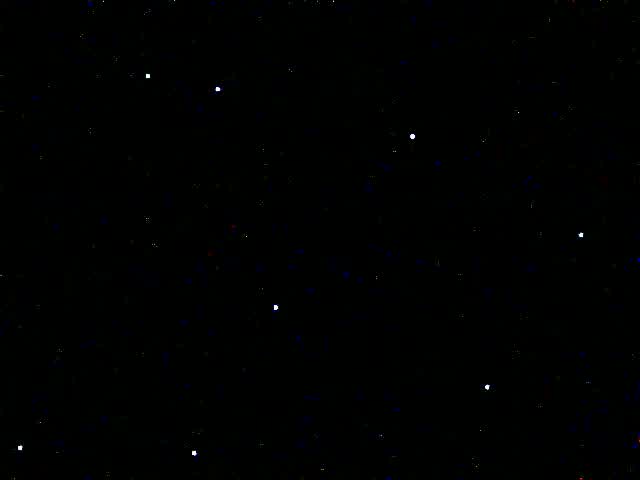}
\caption{Typical video frame showing alpha particle hits. This frame was taken with source-CCD at zero degrees. The frame period is 160~ms. Note the background activity due to noise generated by the CCD}
\label{spots}
\end{figure}

$\alpha$-particles from sources, due to their relatively low energy and limited range, require a clear path to the CCD’s sensitive surface. Most CCDs are covered by a protective glass layer, which is sufficient to stop these particles entirely. To enable detection, this protective layer was carefully removed by softening the adhesive with a micro-torch. A 640×480-pixel CCD sensor with dimensions of $2.88 \times 2.16~\text{mm}^2$ and a 4:3 aspect ratio was selected for this setup. When alpha particles strike the exposed sensor, they deposit charge in small, circular regions that often saturate the affected pixels—producing distinctive, visible signatures, as shown in Figure \ref{spots}.

For detection to occur, alpha particles must reach the CCD’s active region and deposit charge there. However, most semiconductor devices include structural layers above the sensing area that can obstruct low-energy particles. For reference, a 5.5 MeV $\alpha$-particle travels only about $27 \mu m$ in silicon. Because of this short range, the alpha source must be positioned very close to the sensor surface when operating in air, as alpha particles lose energy rapidly with distance. Alternatively, reducing air pressure around the detector, i.e., creating a partial vacuum, can mitigate this limitation by minimizing energy loss in transit.

Another key consideration when using CCDs for particle detection is dead time during data acquisition. Video capture typically occurs at 24 or 30 frames per second (fps), with each frame composed of an exposure phase followed by a readout phase. At 30 fps, a full frame lasts about 33 milliseconds, with readout alone taking up to 20 milliseconds depending on the CCD’s size and design. This results in a potential dead time of up to 75\%, during which new events cannot be recorded. Dead time becomes particularly significant when detecting rare events, such as those occurring at large scattering angles.

Furthermore, many webcams use automatic exposure settings that adjust frame rate based on lighting conditions. In this experiment, data were collected in a dark environment, prompting the camera to lower its frame rate to 6 fps. This reduction extended the frame duration and improved the effective duty cycle, lowering dead time to approximately 17\%—a more favorable condition for event detection. For setups that require finer control of frame rates, open-source software is available to override automatic settings and improve timing precision.

\subsection{$\alpha$-particle source}

The $\alpha$-particle source is a small $^{ 241}Am$ (Americium) with 1 micro-Curie intensity. This source was from an old smoke detector. The extraction of the source was done by specialized personnel as it presents health hazards. The $^{241}Am$ source has a half-life of 432 years, thus a 20 year old source would have lost 5\% of its original intensity. The source is secured in the source holder. The source itself is a small disk that emits $\alpha$ particles in all directions. 

In front of the source, a simple $3~mm$ brass tubing is place to collimate particles. This choice was made for simplicity, but given its short length and wide aperture, the resulting beam is not collinear, but it does have "wings". While most of the particles are in the center,  the lateral spread is such that even at $10^{\circ}$, particles will reach the CCD. Brass is a composite of copper and zinc which will scatter alpha particles. thus a different geometry than a tube would collimate the beam better.

\subsection{The gold foil}

The gold foil is 24 karat gold leaf used for gilding and was chosen because of its purity that is usually 99\% or higher\cite{goldleaf}. Gold leaves with low karat are composite and have less gold content. The average foil thickness according to the manufacturer information is  $0.5 \mu m$.  In comparison, Geiger and Mardsen used a gold foil that was $0.4 \mu m$ thick. These are extremely thin gold foils, approximately 1000 atoms thick, and should be handled with care.

The foil is carefully mounted on a frame or target holder. To facilitate the placement, the foil is cut to size and floated on water and carefully “fished” with the frame. This procedure is reproducible and the same technique is used for the fabrication of targets in accelerators. The foil is very thin, nearly transparent, and should be treated carefully. 

\subsection{The apparatus}

The experimental setup was designed with accessibility and simplicity in mind, using materials and tools readily available to hobbyists or educators. At the core of the design is a transparent, rigid plastic food container—easily found in supermarkets and houseware stores—which serves as the vacuum chamber\cite{rubbermaid}. While many types were considered, the selected container offered a sufficiently tight seal to maintain vacuum during measurements. The seal was further improved by applying a thin layer of petroleum jelly (Vaseline) to the rubber gasket.

To fabricate the structural components, a 3D printer was used. Although alternatives such as laser cutting or milling were considered, 3D printing provided the most straightforward and flexible method for producing all required parts. The components were printed in PLA at 0.2 mm resolution. While effective, this resolution left some surfaces—especially rotating elements—a bit rough, requiring light sanding for smoother motion and better fit. This setup should be viewed as a proof-of-concept, with clear opportunities for refinement and optimization.

Rotation of the alpha source was achieved using a magnetic coupled system. The source is mounted on a rotating platform inside the container, driven externally by neodymium magnets. Two magnets are fixed to the base of the rotating assembly, and external magnets are used to control its position. Although this method saves space and avoids the need for mechanical penetrations, it can be a bit tricky to position the source precisely at specific angles. To aid alignment, angle markers—also 3D printed—are spaced at 10-degree intervals around the rotation axis.

The container includes two custom penetrations: one for the vacuum connection and the other for the CCD signal. The vacuum port consists of a small tube epoxied into the side wall, while the camera signal port includes four embedded pins, also epoxied in place. The USB wires are soldered both inside and outside the chamber, allowing the CCD to transmit data without apparent signal degradation.

A simple hand-operated vacuum pump was used to evacuate the chamber. Although a variety of electrical vacuum pumps could serve this purpose—including rechargeable models—the hand pump was sufficient for the task and readily available. It could draw the pressure down to 0 mm Hg (as indicated on the gauge) in approximately 5–10 minutes, and its manual operation eliminated the need for additional valves or power sources.

\section{Range of alpha particles in air}

The distance between the source and where alpha particles stop is known as the particle range. It is well determined that Americium alpha particles, 5.5 MeV, have a range of about 4 cm at sea level. As the air pressure is lowered, particles will travel further. 

\begin{figure}[h!]
\centering
\includegraphics[width=5in]{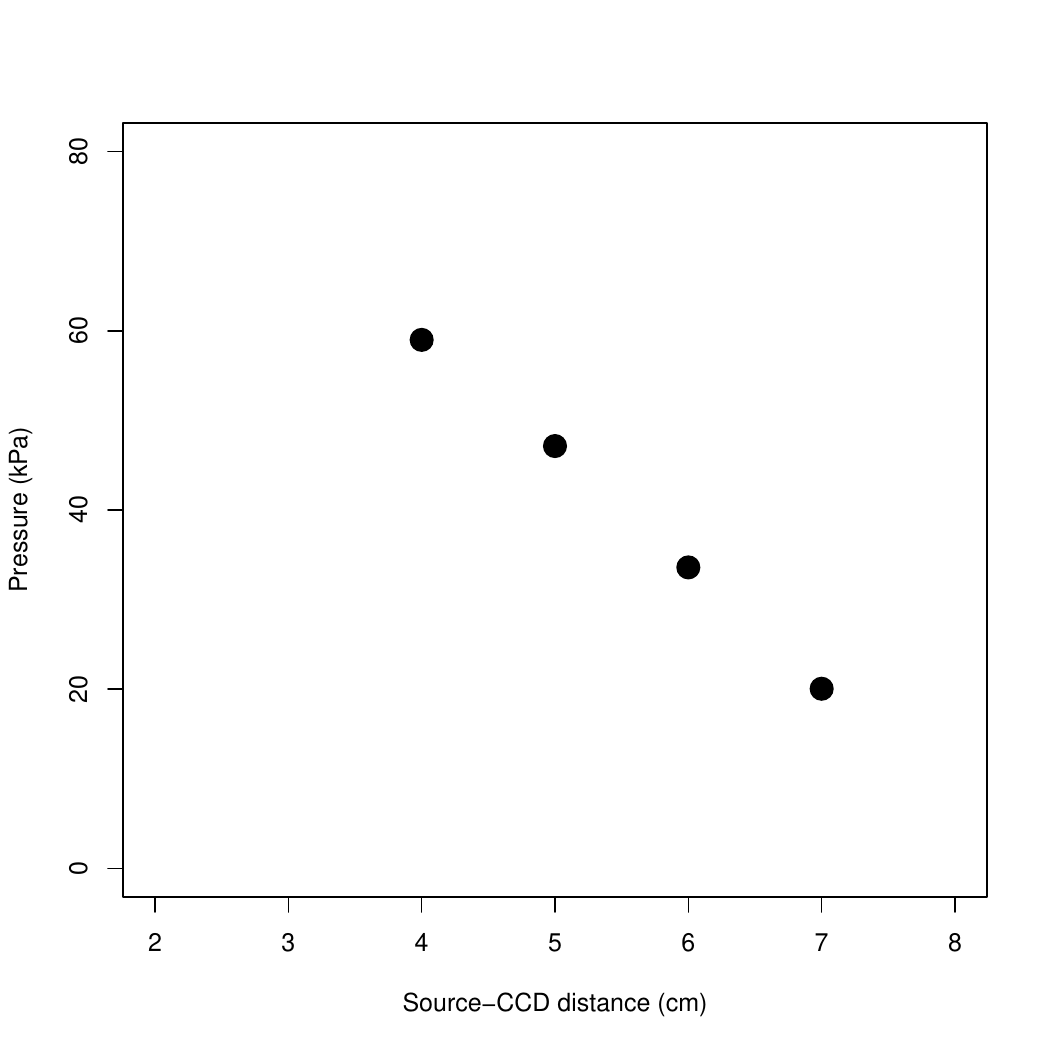}

\caption{The range of alpha particles as a function of chamber pressure. Note at 4 cm, the pressure is already below the atmospheric pressure. Alpha particles need to overcome air and the non
sensitive layers in the CCD.}
\label{range}
\end{figure}

The chamber has provision to place the detector at different distances from the source. 
By varying the distance one can measure the pressure necessary for the alphas to reach the CCD. The procedure followed was to pump down the chamber until particles were observed in the video. Figure \ref{range} shows the plot of the pressure as a function of distance for four points between 4 and 7 cm. 

Note that for 4 cm we already need to pump down the chamber to 60 kPa, which seems to contradict the expectation. However, the alpha particles must penetrate the CCD and charge deposited in the sensitive layers. Thus only when particles have enough energy left will generate a signal in the CCD. For the measurements we have pumped down the chamber as low as possible, and near 0 kPa.

\section{Data Collection and Analysis}

Given the constraints of our apparatus—a broad source collimator, a small CCD sensor, and short source-to-detector distances—gold-foil scattering is most clearly revealed through differential measurements. Data collected with the foil in place are subtracted from data acquired under identical conditions without the foil, suppressing direct (unscattered) counts and isolating those produced by scattering. In foil-free runs, events are detected only within $10^{\circ}$ of the beam axis; thus, any counts recorded at larger angles serve as compelling, though angularly imprecise, evidence of scattering from the gold foil. Note that the $2.88\times 2.16~\text{mm}^2$ CCD spans $\pm2.1^{\circ}$, further limiting angular resolution.

Acquisition times varied from 30 s at $0^{\circ}$ to 7 min at $40^{\circ}$. Videos were captured with QuickTime Player (bundled with macOS), though any standard recording software is suitable. After collecting sufficient statistics, individual frames were extracted using widely available free utilities and web-based tools.

\begin{figure}[h!]
\centering
\includegraphics[width=5in]{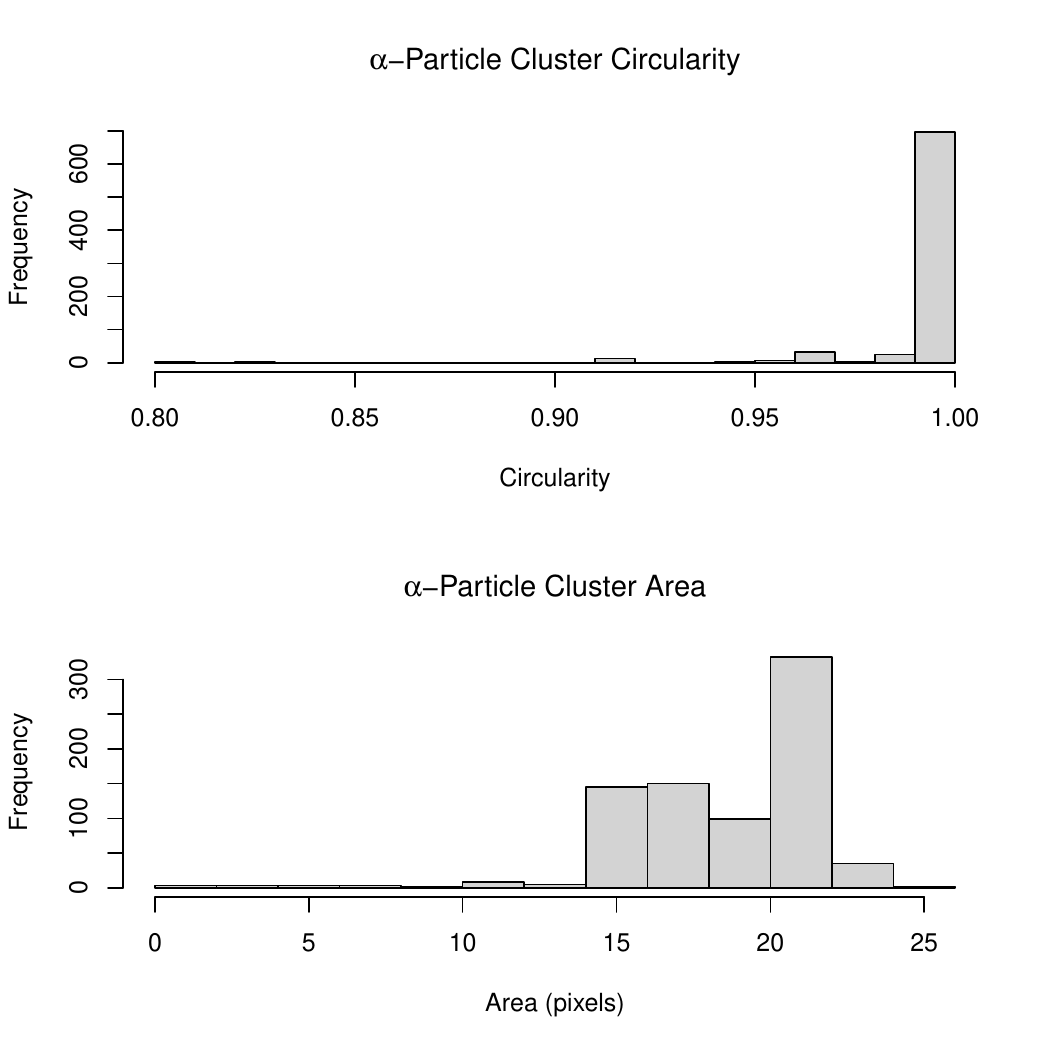}
\caption{$\alpha$-particle cluster properties from the image analysis. (top) The circularity of the clusters and (bottom) the cluster area in pixels. The area reflects the number of pixels where the $\alpha$-particle deposited energy. }
\label{properties}
\end{figure}

Frame analysis was performed in ImageJ\cite{imagej}. Each candidate cluster was characterized by its $(x,y)$ centroid, area, and circularity. Because the camera operated in darkness, the raw frames exhibited significant electronic noise; modest ambient illumination can reduce this noise, yet its attributions differs markedly from genuine $\alpha$ events and is readily filtered. Given that $\alpha$-particles saturate the CCD pixels (intensity 255), a threshold of 100 effectively isolated valid clusters. The ImageJ macro used for batch processing appears in Appendix A, and its CSV output was processed with \texttt{R} scripts\cite{rproject}.  Figure \ref{properties} depicts the cluster circularity and area obtained from the image analysis. For this analysis a cluster with circularity larger than 0.98 and area larger than 14 pixels were required.  Comparison with manual counts showed $100\%$ agreement for these requirements. 

\begin{figure}[h!]
\centering
\includegraphics[width=5in]{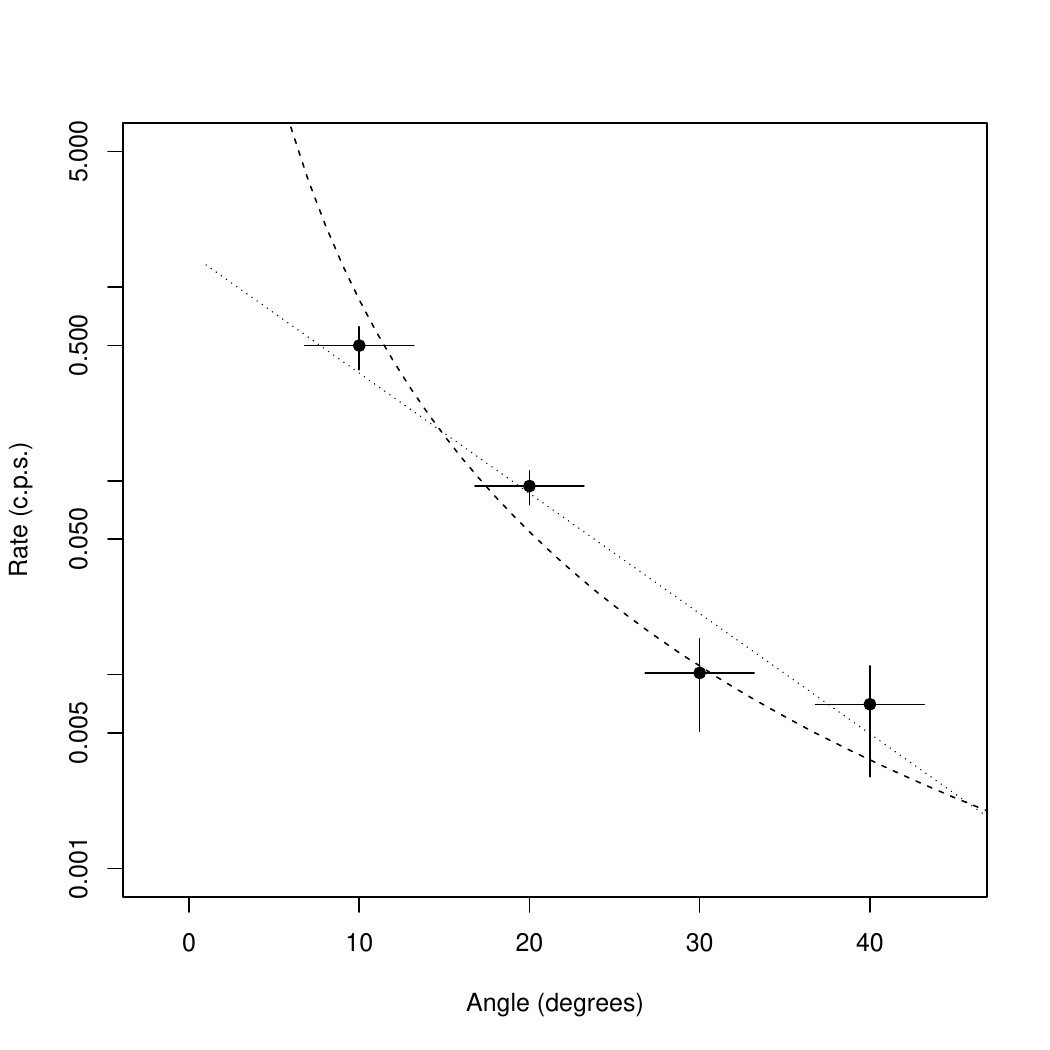}
\caption{Event rate as a function of scattering angle between $10^{\circ}$ and $40^{\circ}$. Each data point represents the difference in counts between measurements taken with and without the gold foil; a significant difference is observed only at $10^{\circ}$. Vertical error bars represent statistical uncertainties based on the number of detected events, while horizontal error bars account for angular uncertainties due to the CCD’s finite size and the estimated positional accuracy of the source. At $30^\circ$, four events were recorded; at $40^\circ$, only one. The dashed curve corresponds to a best-fit of the theoretical Rutherford scattering distribution, $1/\sin^4(\theta/2)$, and the dotted curve shows a normalized exponential model, $e^{-\theta/7.5}$. The current dataset does not allow discrimination between these two functional forms.}
\label{scattering}
\end{figure}

For each angle, cluster counts were normalized to the live acquisition time, assuming constant source activity. Rates at $10^{\circ}, 20^{\circ}, 30^{\circ}$, and $40^{\circ}$ are plotted in Fig.~\ref{scattering}. Horizontal error bars reflect the convolution of the nominal angular bin with CCD acceptance, while vertical bars denote statistical uncertainties. Two curves are overlaid: the dashed line shows the Rutherford functional prediction $1/\sin^{4}(\theta/2)$, and the dotted line shows an exponential $\exp(-\theta/7.5)$, both scaled to best fit the data. With the present statistics, these data cannot discriminate between the models; nevertheless, the observation of $\alpha$-particles beyond $10^{\circ}$ confirms substantial large-angle scattering. The ratio of counts $N(30^{\circ})/N(10^{\circ})$ is 0.02 in this work, as compared to the expected 0.013 from calculations.  For reference, Geiger and Marsden reported a $N(30^{\circ})/N(15^{\circ})$ ratio of 0.06. Thus within experimental uncertainties, the values all fall within the same order of magnitude. These findings strongly suggest that the detected large-angle events arise from $\alpha$ scattering in the gold foil, though higher-precision measurements will be required for a definitive conclusion.

\section{Discussion and Conclusions}
This paper presents a compact and cost-effective nuclear scattering apparatus constructed entirely from commercially available components. At the core of the system is a repurposed CCD sensor from a standard webcam, adapted to function as a particle detector and providing a highly visual experience. The $\alpha$-particle source is salvaged from a smoke detector, while the scattering target—a thin gold foil—is sourced from an art supply store. The supporting structure is 3D-printed in PLA and housed within a rigid plastic food container, modified to serve as a vacuum chamber.

A common question is why $\alpha$-particles scattered from a gold foil can be observed using an $^{241}\mathrm{Am}$ source. The answer lies in the exceptionally large scattering cross section, which is 1755.9 barns. Despite the low probability of wide-angle deflections, the cross section is sufficient to produce detectable rates. For instance, the calculated fraction of $\alpha$-particles scattered at angles greater than $20^\circ$ is $2.6 \times 10^{-4}$. Given the geometry of the setup used in this experiment, this corresponds to approximately three detectable scattering events every five minutes—consistent with observed rates. Specifically, four events were recorded in six minutes at $30^\circ$, and one event was observed over a seven-minute interval at $40^\circ$.

Designed with accessibility and ease of replication in mind, the apparatus is suitable for both classroom demonstrations and hands-on experimental activities. While fully functional, the system offers several avenues for enhancement. For instance, a more effective collimator—such as a multi-aperture design made from small washers—could help generate a more parallel $\alpha$-particle beam. Similarly, equipping the CCD with a collimator would improve angular resolution. The overall mechanical precision could also be improved through the use of higher-resolution 3D printers or laser-cut components, resulting in better alignment and structural integrity.

Despite its simplicity, the apparatus yields promising results. It successfully detects $\alpha$-particles scattered at large angles—one of the principal goals of this project. While the event rate is relatively low, techniques such as motion-triggered acquisition could allow for longer integration times without generating unnecessary data. Additionally, using slightly thicker gold foils may improve detection efficiency while maintaining angular sensitivity.

In summary, this work introduces a low-cost, modular nuclear scattering apparatus with significant educational potential. With continued refinement, it could serve as an effective platform for teaching core principles of nuclear physics and experimental techniques. It is equally well-suited for public demonstrations, offering an engaging and tangible window into the atomic world. Looking ahead, the setup could be adapted to detect protons via the $^{14}\text{N}(\alpha, p)$ reaction, expanding its functionality. Protons, having roughly one-fourth the ionizing power of $\alpha$-particles and a longer range, would appear as smaller spots on the CCD. Furthermore, experimenting with alternative foil materials could provide deeper insights into scattering behavior and nuclear interactions.

\appendix*   % Omit the * if there's more than one appendix.

\section{ImageJ Macro for Image Processing}

ImageJ macro used to process images. This macro will produce a CSV file containing the cluster position (X, Y), area, and circularity. Both coordinates are in pixels. 

\begin{verbatim}
// Choose the input folder where all images for one angle are.
//
dir = getDirectory("Choose a folder");
list = getFileList(dir);
outputFile = dir + "Particle_Results.csv";

// Create results file with header
File.delete(outputFile);
File.append("Filename,X,Y,Area,Circularity\n", outputFile);

// Set threshold range here:
thresholdMin = 100;
thresholdMax = 255;

setBatchMode(true);

for (i = 0; i < list.length; i++) {
    name = list[i];
    if (endsWith(name, ".tif") || endsWith(name, ".jpg") || endsWith(name, ".png")) {
        open(dir + name);

        // Step 1: Convert to grayscale (8-bit)
        run("8-bit");

        // Step 2: Smooth
        run("Smooth");

        // Step 3: Manual thresholding
        setThreshold(thresholdMin, thresholdMax);
        setOption("BlackBackground", false);
        run("Convert to Mask");

        // Step 4: Analyze Particles
        run("Set Measurements...", "area centroid circularity redirect=None decimal=3");
        run("Analyze Particles...", "size=0-Infinity circularity=0.00-1.00 show=Nothing clear include");

        // Step 5: Save particle data
        n = nResults;
        for (j = 0; j < n; j++) {
            x = getResult("X", j);
            y = getResult("Y", j);
            area = getResult("Area", j);
            circ = getResult("Circ.", j);
            File.append(name + "," + x + "," + y + "," + area + "," + circ + "\n", outputFile);
        }

        run("Clear Results");
        close();
    }
}

setBatchMode(false);


\end{verbatim}

\begin{acknowledgments}

This work was partially developed as a course project in PHY-579 (Special Topics) offered by the Department of Physics and Astronomy at Stony Brook University, and further advanced during the QuarkNet Summer Program at Brookhaven National Laboratory.

\end{acknowledgments}

\end{document}